\documentclass[11pt]{article}
\usepackage{graphicx}
\usepackage{amssymb}
\usepackage{amsmath}
\usepackage[hidelinks]{hyperref}
\usepackage{booktabs}
\usepackage{subcaption}
\usepackage{subcaption}
\usepackage{tikz}
\usepackage{url}
\renewcommand{\title}[1]{{\noindent\large\bfseries#1\medskip\\}}
\renewcommand{\author}[2]{{\noindent #1 \medskip\\ \small #2 \medskip\\}}
\usepackage[letterpaper,margin=20mm]{geometry}
\usepackage[style=apa, backend=biber,sorting=none,labelnumber]{biblatex}
\usepackage{csquotes}
\usepackage{svg}
\usepackage{xpatch} 

\begin{document}

\title{Latent Spatial Heterogeneity in U.S. Cancer Mortality: A Multi-Site Clustering and Spatial Autocorrelation Analysis.}

\author{Emmanuel Kubuafor\textsuperscript{1*}, Dennis Baidoo\textsuperscript{1}, Agnes Duah\textsuperscript{2} , Robert Amevor\textsuperscript{3}, Onyedikachi Joshua Okeke\textsuperscript{4}, Dorcas Quaye\textsuperscript{5},
Peter Ofori Appiah\textsuperscript{6}
}
{
\textsuperscript{1} Department of Mathematics and Statistics, University of New Mexico, Albuquerque, NM, USA. \\ 
\textsuperscript{2} Department of Biostatistics, University of Indiana-Bloomington, Bloomington, IN, USA. \\ 
\textsuperscript{3} Arnold School of Public Health, University of South Carolina, SC, Columbia, USA. \\ 
\textsuperscript{4} Department of Geography and Environmental Studies, Texas State University, San Marcos, TX, USA. \\ 
\textsuperscript{5} Department of Anatomy, University of Ghana, Accra, Ghana.\\
\textsuperscript{6} Department of Medical Microbiology, University of Ghana, Accra, Ghana. \\

*Corresponding author: \texttt{ekubuafor28@gmail.com}
}

\noindent \textbf{{\Large Abstract}}\\

\noindent Cancer mortality continues to vary significantly across U.S. regions, highlighting a major public health challenge.
While national cancer rates have declined over recent decades, regional patterns persist, reflecting underlying socioeconomic, behavioral, and environmental risk factors.

\noindent This research set out to explore and delineate spatial patterns and mortality distributions for various cancer types across U.S. states between 1999 and 2021. The aim was to uncover region-specific cancer burdens and inform geographically targeted prevention efforts. We analyzed state-level cancer mortality records sourced from the CDC WONDER platform, concentrating on cancer sites consistently reported across the 48 contiguous states and Washington, D.C., excluding Hawaii, Alaska, and Puerto Rico. Multivariate clustering using Mahalanobis distance grouped states according to similarities in mortality profiles. Spatial autocorrelation was examined for each cancer type using both Global Moran’s I and Local Indicators of Spatial Association (LISA). Additionally, the Getis-Ord \( G_i^* \) statistic was applied to detect cancer-specific hotspots and cold spots.


\noindent Clustering revealed three state groups with distinct mortality profiles. One cluster, concentrated in the South and Appalachia, exhibited elevated mortality rates for liver, pancreatic, colorectal, and breast cancers. Global spatial autocorrelation analysis using Moran’s I revealed statistically significant clustering across all examined cancer types. The most pronounced spatial dependencies were identified for kidney, liver, and lung cancers. LISA maps identified lung cancer hotspots in southeastern states such as Kentucky, Arkansas, and Tennessee, and cold spots in western states including Utah and Colorado. A hotspot frequency analysis further identified Virginia, Missouri, and Maryland as persistently high-burden states.

\noindent Cancer mortality in the United States exhibits strong and statistically significant spatial clustering. These patterns highlight persistent regional disparities and support the need for geographically targeted cancer prevention and control strategies.


\section{\Large Introduction}

\noindent Cancer-related death rates have generally declined across the United States; however, regional inequalities remain. Persistent gaps in health outcomes are especially evident in the South, the Appalachian region, and portions of the Midwest, where multiple cancer types show consistently elevated mortality rates \parencite{Henley2017, Singh2017}. These patterns are shaped by a multifaceted interaction of structural and socioeconomic determinants, including poverty, limited access to high-quality healthcare, and regionally elevated levels of behavioral risk exposures such as tobacco use \parencite{DeSantis2019}.

\noindent \textcite{Sabharwal2024} emphasizes the widening gap in cancer mortality between Appalachian and non-Appalachian regions, highlighting persistent socioeconomic and healthcare access barriers that contribute to worse outcomes in vulnerable populations. Addressing these spatial disparities is critical for designing targeted interventions that tackle the root causes of inequity and improve cancer outcomes in high-burden areas. Racial and ethnic inequalities further exacerbate geographic disparities, as minority populations frequently experience disproportionately higher mortality due to systemic barriers to care \parencite{Wang2022}.\\

\noindent Spatial epidemiology provides a valuable approach for uncovering geographic variation in disease patterns and interpreting cancer mortality disparities. Tools like Moran’s I, Local Indicators of Spatial Association (LISA), and the Getis-Ord $G_i^*$ statistic are frequently applied to detect areas with significantly elevated or reduced mortality, identifying spatial hotspots and cold spots \parencite{Anselin1995, Waller2004}.

\noindent For example, \textcite{Schulz2022} applied spatial clustering and regression models to analyze breast cancer mortality at the county level, identifying strong spatial patterns shaped by local sociodemographic factors. Bayesian spatial modeling approaches further refine this analysis by providing smoothed risk estimates, particularly in areas with sparse data, and enabling the simultaneous modeling of multiple cancer types \parencite{Gao2019, Cramb2015}. These models account for spatial dependence and uncertainty, offering more precise estimates of cancer risk. Additionally, advanced methods, including spatial scan statistics and Multivariate Conditional Autoregressive (MCAR) models, are capable of detecting spatial groupings that are irregular in shape and involve multiple diseases, effectively capturing complex spatial configurations that conventional techniques may miss \parencite{Yin2018, Prates2022}.


\noindent Recent advances have incorporated machine learning techniques, especially unsupervised clustering methods, into cancer epidemiology to reveal latent patterns that may not be captured by traditional spatial analyses. \textcite{Amin2019} employed spatial clustering techniques to group U.S. counties based on breast cancer mortality and incidence, revealing regional patterns associated with sociodemographic and environmental risk factors. Similarly, \textcite{Moore2017} used spatial clustering to identify lung cancer mortality patterns across U.S. counties, revealing clusters aligned with socioeconomic disadvantage and environmental exposures such as air pollution. Beyond K-means clustering, machine learning methods such as hierarchical clustering, random forests, and deep learning are increasingly integrated with spatial epidemiology to enhance predictive accuracy and uncover complex interactions \parencite{Wang2019, Lim2025}. These integrative approaches provide a powerful framework for tailoring cancer control strategies to regional needs and improving the equity and efficiency of public health interventions.\\

\noindent Multiple upstream factors, such as income inequality, educational opportunities, healthcare availability, and environmental stressors, are widely recognized as key influences on disparities in cancer burden. \textcite{Moss2021} found that socioeconomic disadvantage, lack of screening services, and lifestyle factors such as obesity significantly contribute to spatial variation in cancer mortality across U.S. counties. Similarly, \textcite{Anderson2023} identified obesity rates and mammography utilization as key predictors of breast cancer mortality, with their impacts varying significantly across regions. \textcite{Cheng2021}  highlighted how localized socioeconomic disadvantage can shape both cancer diagnosis rates and patient survival outcomes. In addition, \textcite{Bevel2023}  demonstrated that broader contextual exposures, such as environmental degradation, including air quality issues, and inadequate access to nutritious food, also increase the risk of cancer. Moreover, socioeconomic disadvantage can delay diagnosis and limit access to treatment, contributing to poorer survival outcomes \parencite{Bourgeois2024}.\\

\noindent Traditional spatial studies often focus on individual cancer types, which can obscure broader spatial patterns of cancer burden. Multivariate spatial modeling approaches allow for the joint analysis of multiple cancer types, facilitating the identification of regions with overlapping risks where coordinated interventions may be most effective. \textcite{Vicente2022} applied multivariate Bayesian disease mapping to identify regions with simultaneously elevated mortality for several cancers, providing evidence for integrated public health responses. Similarly, \textcite{Dong2022} used multiscale spatial modeling, incorporating geographic random forests to identify cancer mortality determinants across counties, revealing intricate spatial interactions among cancers and their risk factors. Tools such as spatial scan statistics and MCAR models further enhance the capacity to detect irregularly shaped and multivariate disease clusters, offering a more comprehensive spatial representation of cancer risk \parencite{Lin2016}.\\

\noindent Despite these methodological advances, many studies remain limited in scope, often examining only one cancer type or focusing on a narrow geographic scale. There remains a critical need for comprehensive analyses that encompass multiple cancer types across the entire U.S. landscape. To address this gap, our study analyzes state-level mortality data for 16 leading cancer types from 1999 to 2021 across 49 U.S. states, including the 48 contiguous states and Washington, D.C., while excluding Hawaii, Alaska, and Puerto Rico. We employed a hybrid methodology that combines hierarchical clustering with spatial diagnostics such as Moran’s $I$, LISA, and the Getis-Ord $G_i^*$ statistic to uncover recurrent spatial clusters and categorize states exhibiting comparable mortality patterns. This integrative framework improves our understanding of regional disparities in cancer outcomes and informs the design of targeted, data-informed public health interventions tailored to specific geographic contexts.

\section{\Large Methodology}

\subsection{Data Source and Inclusion Criteria}

\noindent We analyzed cancer mortality data from the CDC WONDER United States Cancer Statistics database for the years 1999 to 2021. The dataset includes state-level, age-adjusted death rates per 100,000 population for leading cancer sites, standardized to the 2000 U.S. population. Only cancer types with non-suppressed, consistently reported data across all states and years were included. Records with fewer than 16 deaths or zero counts were excluded in accordance with CDC privacy standards. Puerto Rico, Alaska, Hawaii, and other U.S. territories were excluded to maintain geographic continuity and comparability in the spatial analysis. Population estimates for 2005 exclude individuals displaced by Hurricanes Katrina and Rita. Sixteen leading cancer sites met inclusion criteria: lung and bronchus, colorectal, pancreas, breast, prostate, liver, leukemia, non-Hodgkin lymphoma, esophagus, stomach, kidney and renal pelvis, urinary bladder, brain and other nervous system, ovary, oral cavity and pharynx, and myeloma.

\subsection{Age Adjustment and Rate Standardization}
\noindent Age-adjusted mortality rates in the data were estimated using the direct standardization technique, referencing the 2000 U.S. standard population. This method helps control for confounding introduced by variation in age distributions across states and time periods. The resulting rate, expressed per 100{,}000 individuals, is computed as:

\begin{align*}{\tag 1}
\text{Age-adjusted rate } = \sum_{i=1}^{k} \left( \frac{D_i}{P_i} \right) w_i \times 100{,}000
\end{align*}

\noindent where $D_i$ denotes the observed number of deaths in age group $i$, $P_i$ is the corresponding population size, and $w_i$ represents the weight assigned to that age group based on its proportion in the reference population.




\noindent The average age-adjusted mortality rate for each cancer site within each state over the 23-year period was arranged in a state-by-cancer matrix where rows represented states and columns represented cancer sites. The resulting matrix was mean-centered and scaled by standard deviation (z-score normalization) to eliminate scale-induced bias.

\subsection{Unsupervised Hierarchical Clustering Using Mahalanobis Distance}

\noindent To identify groups of U.S. states with similar cancer mortality profiles, we applied hierarchical agglomerative clustering to the standardized matrix of average age-adjusted mortality rates for the 16 leading cancer sites consistently reported across all 49 U.S states (including D.C) considered in this study. Each state was represented by a 16-dimensional vector, with features standardized using z-scores to allow for meaningful comparison.

\noindent Unlike Euclidean distance, which assumes independence and equal scale across variables, we used Mahalanobis distance to account for the covariance structure among cancer sites. Given two states with standardized mortality vectors $\mathbf{x}$ and $\mathbf{y}$, and a covariance matrix $\Sigma$, Mahalanobis distance ($D_{M}(\mathbf{x}, \mathbf{y})$) is defined as:

\begin{align*}{\tag 2}
D_M(\mathbf{x}, \mathbf{y}) = \sqrt{(\mathbf{x} - \mathbf{y})^\top \Sigma^{-1} (\mathbf{x} - \mathbf{y})}.
\end{align*}

\noindent Using the Mahalanobis distance matrix, we applied hierarchical clustering with Ward’s method to group states into compact clusters based on their multivariate cancer mortality profiles. The silhouette method was then used to determine the optimal number of clusters, selecting the configuration that maximized overall cluster cohesion and separation. 

\noindent Cluster assignments were then joined with U.S. state geographic shapefiles and visualized through choropleth mapping to explore geographic continuity, spatial structure, and potential regional disparities in cancer mortality. These clusters provide a high-level summary of cancer burden similarity across states and support further investigation into shared structural or environmental determinants.

\subsection{Spatial Autocorrelation: Global Moran's I}
\noindent To explore spatial dependencies in cancer mortality across the United States, we implemented the Global Moran’s \( I \) statistic independently for each cancer site. The analyses were conducted using  U.S. shapefiles for the 48 states and the District of Columbia, excluding Alaska, Hawaii, and U.S. territories to minimize edge effects. This statistic measures the overall degree of spatial autocorrelation, capturing whether states with similar mortality rates tend to be geographically clustered across the entire study area. High positive values of  \( I \) (approaching +1) indicate strong spatial clustering, where neighboring states share similarly high or low mortality rates. Negative values (approaching -1) suggest spatial dispersion, while values near zero reflect spatial randomness.

\noindent Let \( \lambda_i \) denote the age-adjusted mortality rate for a specific cancer site in state \( i \), and \( \bar{\lambda} \) the mean rate for that cancer site across all \( N = 49 \) jurisdictions (including D.C.). The spatial relationship between states \( i \) and \( j \) is captured by the weight \( w_{ij} \), derived using a row-standardized queen contiguity weights matrix, which assigns nonzero weights to neighboring states sharing a border or corner:

\begin{align*}{\tag{3}}    
w_{ij} = 
\begin{cases}
\frac{1}{n_i} & \text{if states } i \text{ and } j \text{ are neighbors} \\\\
0 & \text{otherwise}
\end{cases}
\end{align*}

\noindent where \( n_i \) is the number of neighbors of state \( i \), ensuring that \( \sum_j w_{ij} = 1 \) for all \( i \).

\noindent The Global Moran’s \( I \) is then calculated as:

\begin{align*}{\tag{4}}
I = \frac{N}{\sum_{i=1}^{N} \sum_{j=1}^{N} w_{ij}} \cdot \frac{\sum_{i=1}^{N} \sum_{j=1}^{N} w_{ij}(\lambda_i - \bar{\lambda})(\lambda_j - \bar{\lambda})}{\sum_{i=1}^{N} (\lambda_i - \bar{\lambda})^2}.
\end{align*}

\noindent The numerator measures the weighted spatial covariance in mortality rates across neighboring states, while the denominator captures the overall variance in \( \lambda \). 
Statistical inference was based on the analytical expectation and variance of Moran’s \( I \) under the null hypothesis of spatial randomness. Standardized \( z \)-scores were computed and compared to the normal distribution to determine significance. Cancer sites with \( p \)-values less than \( 0.05 \) were interpreted as exhibiting statistically significant spatial clustering.


\subsection{Local Indicators of Spatial Association (LISA)}

\noindent Although global measures such as Moran’s $I$ can quantify the overall degree of spatial autocorrelation across all states, they provide limited insight into where spatial clustering or local anomalies occur. To uncover these localized patterns in cancer mortality, we employed Local Indicators of Spatial Association (LISA), specifically the local Moran’s $I_i$ statistic, for each of the 16 cancer sites across the 49 states, including D.C.

\noindent For a given state $i$, the local Moran’s statistic is defined as:

\begin{align*}{\tag 5}
I_i = \frac{(\lambda_i - \bar{\lambda})}{s^2} \sum_{j=1}^{n} w_{ij}(\lambda_j - \bar{\lambda}),
\end{align*}

\noindent where $\lambda_i$ denotes the average age-adjusted mortality rate for a specific cancer site in state $i$, $\bar{\lambda}$ is the mean mortality rate for that cancer site across the 49 states (including the District of Columbia), and $w_{ij}$ is the spatial weight between states i and j, based on the row-standardized queen contiguity matrix. The denominator $s^2$ represents the sample variance of the mortality rates $\lambda$, calculated across all states for that cancer type. This standardization ensures comparability across states with differing mean levels and scales.

\noindent The statistic $I_i$ measures how strongly the mortality rate in a given state differs from the national average in relation to its neighbors. Positive values of $I_i$ indicate that a state has similar mortality rates to its neighbors (either high or low), while negative values suggest that a state differs substantially from those around it. Each $I_i$ value was tested for statistical significance by comparing the observed $I_i$ values to their expected values under spatial randomness, assuming a normal distribution. This approach provided $z$-scores for inference. States with $p$-values below 0.05 were considered to exhibit statistically significant local spatial association. These states were then classified into four types: (1) High-High clusters, where high mortality rates are surrounded by similarly high values (hotspots); (2) Low-Low clusters, where low mortality rates are surrounded by low values (cold spots); (3) High-Low outliers, where high mortality rates are bordered by lower-rate neighbors; and (4) Low-High outliers, where low mortality rates are surrounded by higher-rate neighbors.
States with non-significant $I_i$ values were not assigned to any spatial cluster and are interpreted as having no statistically meaningful local pattern.
Cluster maps were created to display the spatial variation, with separate panels for each cancer site and shading used to indicate statistically significant local clustering.



\subsection{Hotspot Detection Using Getis-Ord Gi\texorpdfstring{\textsuperscript{*}}{*}}

\noindent To analyze the spatial concentration of extreme cancer mortality rates, we applied the Getis-Ord $G_i^*$ statistic separately for each of the 16 cancer types. For each cancer site, a $G_i^*$ value was computed for every state considered in this study, identifying localized clusters of significantly high or low mortality relative to the national mean. The statistic is defined as:

\begin{align*}{\tag 6}
G_i^* = \frac{\sum_j w_{ij} \lambda_j - \bar{\lambda} \sum_j w_{ij}}{S \sqrt{\left[ \frac{n \sum_j w_{ij}^2 - \left(\sum_j w_{ij}\right)^2}{n-1} \right]}},
\end{align*}

\noindent where $\lambda_j$ represents the age-adjusted mortality rate for a specific cancer site in neighboring state $j$, $w_{ij}$ is the spatial weight between states $i$ and $j$ derived from a row-standardized queen contiguity matrix, $\bar{\lambda}$ is the mean age-adjusted mortality rate for the cancer site across the $n = 49$ states (including D.C.), and $S$ is the sample standard deviation of $\lambda$.

\noindent The resulting $G_i^*$ values are interpreted as standard normal $z$-scores. Elevated positive values indicate spatial clustering of high mortality (hotspots), while large negative values indicate clustering of low mortality (cold spots). Values near zero suggest no meaningful spatial concentration. 

\noindent Statistical significance was evaluated analytically under the assumption of spatial randomness. States were categorized as significant hotspots or cold spots using a two-sided significance test with a threshold of $\alpha = 0.05$. This procedure was repeated independently for each of the 16 cancer types. Additionally, a frequency analysis was conducted to tally how often each state appeared as a statistically significant hotspot.




\section{\Large Results}

\noindent Between 1999 and 2021, we analyzed 18,618 state-level cancer mortality records spanning 49 U.S. states (including D.C). We studied 16 leading cancer site, analyzing annual observations across each state and cancer site combination.. The age-adjusted mortality rate across all records ranged from 1.10 to 81.3 deaths per 100,000, with a mean of 10.17 and a standard deviation of 11.51. Of the 16 cancer sites consistently reported across the 49 states studied (including D.C), lung and bronchus, colorectal, and breast cancers exhibited the highest average mortality rates of 46.74, 16.28, and 12.38 per 100,000, respectively.

\begin{figure}[t!]
    \centering
    \begin{subfigure}[htbp!]{0.45\linewidth}
        \centering
        \includegraphics[width=\linewidth]{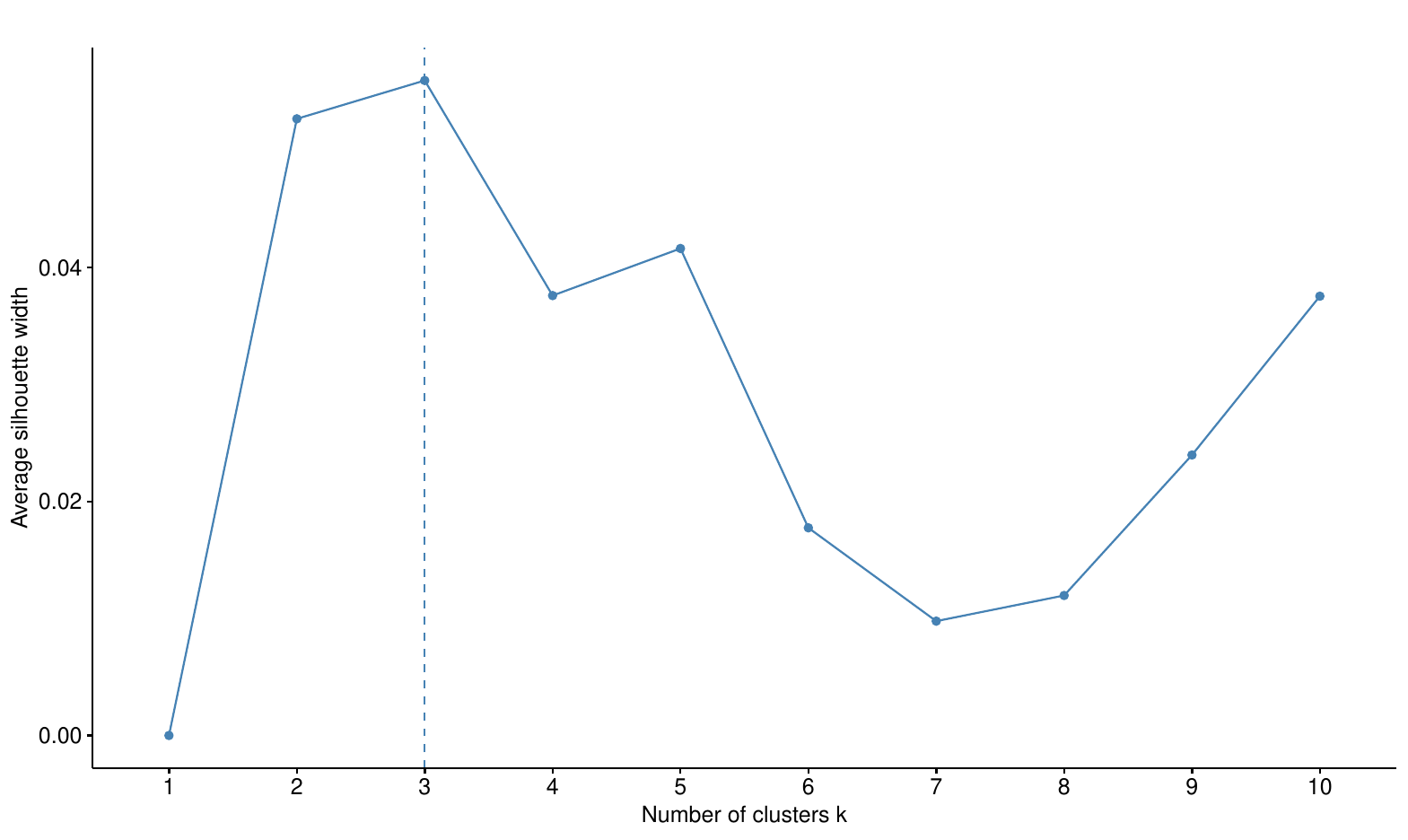}
        \caption{Silhouette plot showing the optimal number of clusters (k = 3).}
        \label{fig:1(a)}
    \end{subfigure}
    \hspace{0.05\linewidth}
    \begin{subfigure}[htbp!]{0.45\linewidth}
        \centering
        \includegraphics[width=\linewidth]{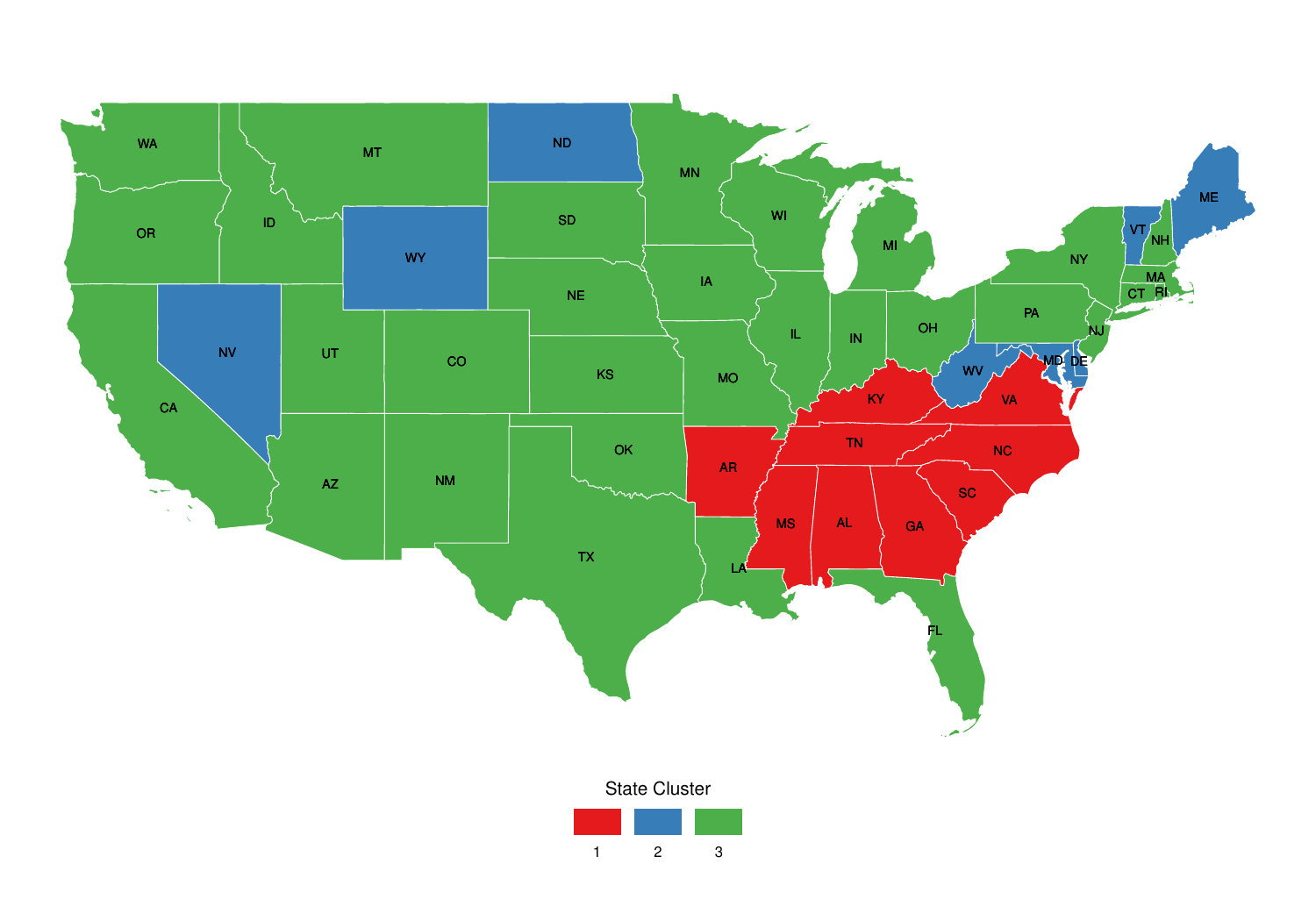}
        \caption{Geographic distribution of cluster assignments.}
        \label{fig:1(b)}
    \end{subfigure}

    \vspace{0.5cm}
    \begin{subfigure}[htbp!]{0.7\linewidth}
        \centering
        \includegraphics[width=\linewidth]{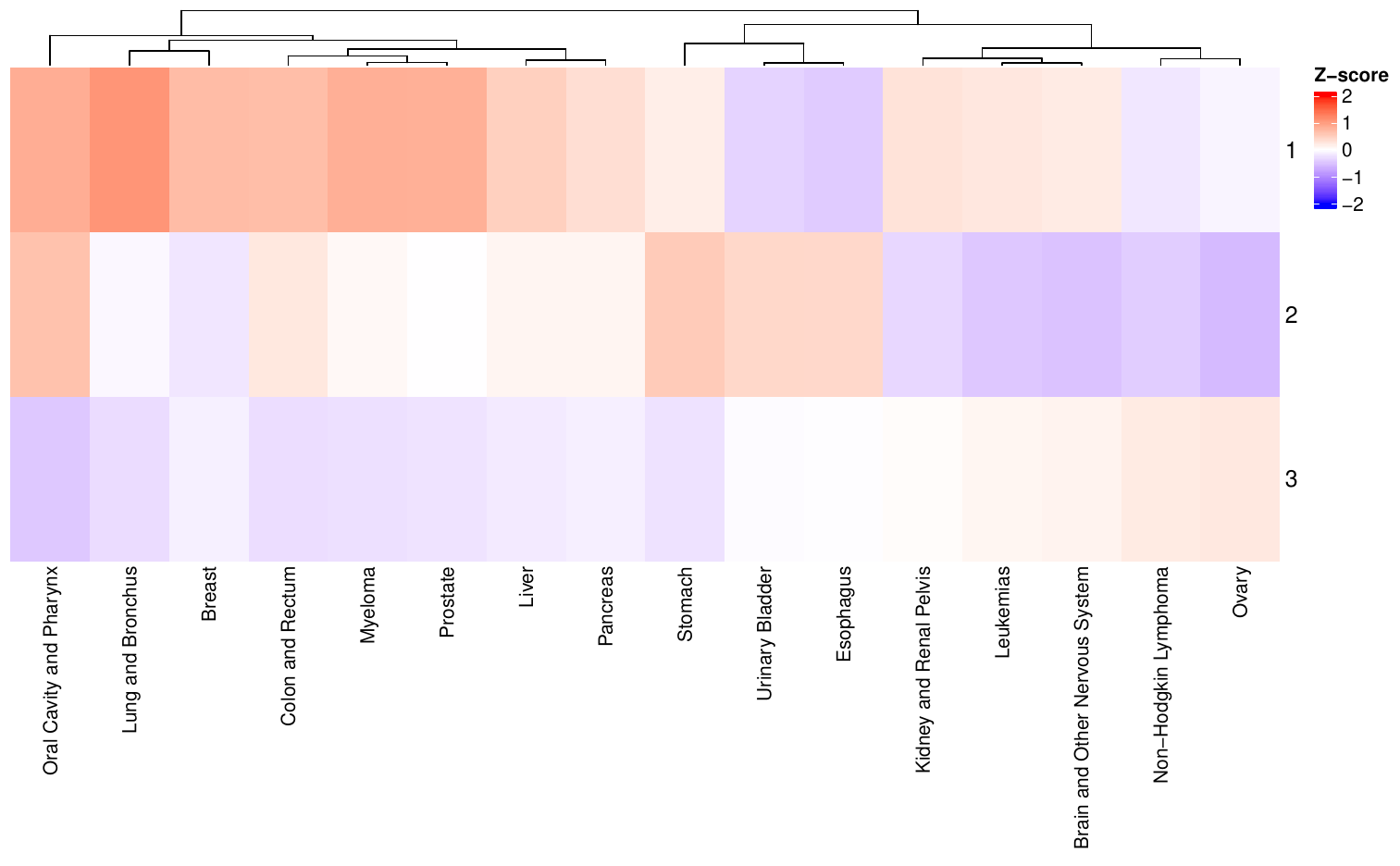}
        \caption{Heatmap of standardized cancer mortality profiles by cluster and cancer type.}
        \label{fig:1(c)}
    \end{subfigure}

    \caption{Clustering results based on hierarchical clustering with Mahalanobis distance: (a) silhouette analysis showing optimal cluster number; (b) spatial distribution of clusters across U.S. states; (c) heatmap of cancer mortality profiles by cluster.}
    \label{fig:clustering_results}
\end{figure}

\noindent Hierarchical clustering based on Mahalanobis distance grouped U.S. states into distinct multivariate cancer mortality profiles. The silhouette method identified a three-cluster solution as optimal, maximizing within-cluster cohesion and between-cluster separation, as indicated by a clear peak at k = 3 in the silhouette plot (Figure~\ref{fig:1(a)}). Figure~\ref{fig:1(b)} displays how cluster assignments are geographically distributed among U.S. states. Cluster 1, which includes Alabama, Kentucky, Mississippi, and Tennessee, is concentrated in the southeastern and Appalachian regions. Cluster 2, comprising states such as California, Arizona, Maine, and North Dakota, is geographically dispersed. Cluster 3 includes the largest number of states, spanning much of the Northeast, Midwest, and West, including New York, Illinois, Minnesota, and Colorado.


\pagebreak
\noindent To further analyze these groupings, Figure~\ref{fig:1(c)} displays the average standardized mortality rates (Z-scores) by cluster and cancer type. Cluster 1 shows consistently elevated mortality across a broad range of cancers, particularly oral cavity and pharynx, lung and bronchus, breast, colon and rectum, myeloma, liver, and prostate. Cluster 2 displays moderate rates, with elevated mortality in stomach and in oral cavity and pharynx cancers. Cluster 3 is characterized by uniformly lower-than-average mortality across nearly all cancer types.\\

\begin{figure}[htbp!]
    \centering

    \begin{subfigure}[b]{0.8\linewidth}
        \centering
        \includegraphics[width=\linewidth]{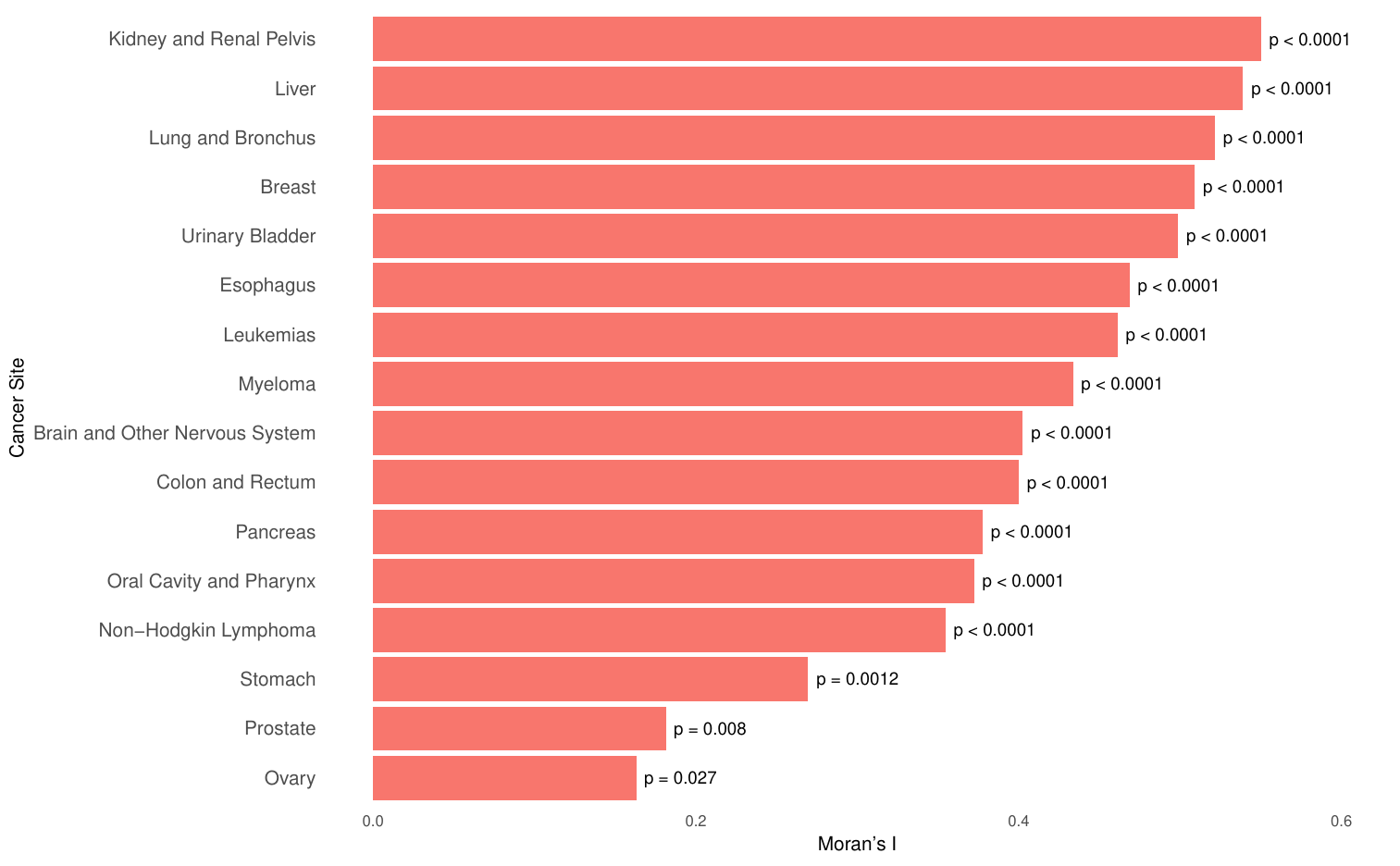}
        \caption{Global Moran’s I values for cancer mortality rates by site.}
        \label{fig:2(a)}
    \end{subfigure}
    \hspace{0.05\linewidth}
    \begin{subfigure}[b]{\linewidth}
        \centering
        \includegraphics[width=\linewidth]{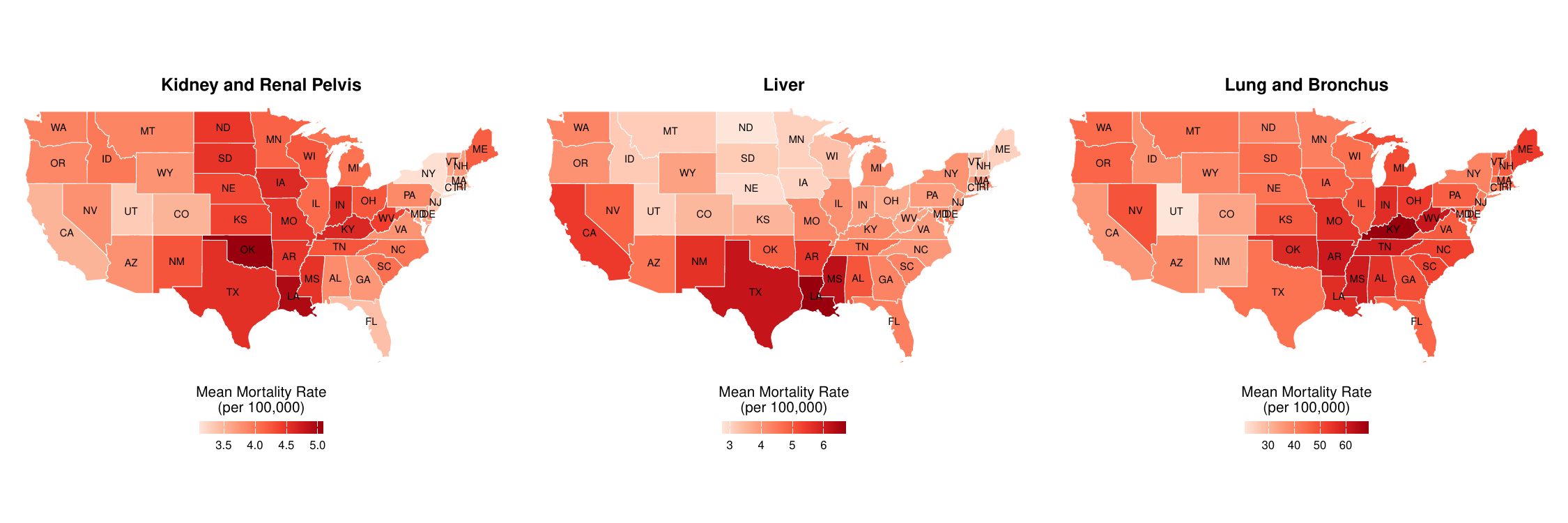}
        \caption{Top three cancer sites with the strongest spatial clustering.}
        \label{fig:2(b)}
    \end{subfigure}
    \caption{Spatial autocorrelation of cancer mortality rates, 1999–2021. (a) Global Moran’s I values across cancer sites. (b) Highest clustering intensity for kidney, liver, and lung cancers.}
    \label{fig:spatial_autocorrelation}
\end{figure}

\noindent To evaluate the spatial concentration of cancer mortality across the U.S., we calculated Global Moran’s I for all 16 uniformly reported cancer sites. The results showed significant spatial autocorrelation ($p < 0.05$) for each cancer type, indicating notable geographic clustering (Figure \ref{fig:2(a)}). The highest levels of spatial autocorrelation were observed for kidney and renal pelvis (Moran’s I = 0.57), liver (0.55), and lung and bronchus (0.53) cancers, all with p-values below 0.0001. Other cancer types, such as breast, bladder, esophagus, and leukemias, also showed relatively strong clustering (Moran’s I $>$ 0.4), while stomach, prostate, and ovary had weaker but still significant spatial patterns, indicating non-random geographic distributions.

\newpage
\noindent To further visualize these patterns, Figure \ref {fig:2(b)} displays state-level maps of average age-adjusted mortality rates for the three cancer types with the strongest spatial clustering based on Global Moran’s I: lung and bronchus, liver, and kidney. Lung cancer showed the highest mortality burden (30–60+ per 100,000), with concentrated rates in the Southeast and Appalachian states such as Kentucky, West Virginia, and Mississippi. Liver cancer mortality (3–6 per 100,000) was elevated along the Gulf Coast, particularly in Texas and Louisiana, and parts of the Southwest. Kidney cancer showed more diffuse patterns (3.5–5 per 100,000), with the highest rates observed in Oklahoma and Louisiana. These patterns highlight distinct regional disparities in cancer burden across the U.S.

\noindent Local Indicators of Spatial Association (LISA) analysis revealed significant spatial clustering of cancer mortality across the U.S. states studied. High-high clusters were most frequently observed in Kentucky and Tennessee, especially for lung and bronchus, colorectal, and breast cancers. Mississippi, Louisiana and Arkansas also appeared as high-high clusters for liver, oral cavity/pharynx and colorectal cancers. Additionally, Missouri and Illinois showed high-high clustering for kidney and renal pelvis, non-Hodgkin lymphoma and leukemia cancers. These patterns highlight concentrated cancer mortality burdens in parts of the South and Appalachian region.


\noindent Low-low clusters, indicating areas with low mortality surrounded by similarly low-rate neighbors, were concentrated in the Southwest and Mountain West. States such as Utah, Arizona, Colorado, and New Mexico frequently exhibited low mortality rates for pancreatic, lung, esophageal, and oral cavity cancers. Wyoming, South Dakota, Montana and Minnesota also show low-low clusters for breast and liver cancers. These regions may reflect more favorable health profiles or greater access to preventive care.

\noindent Several cancers displayed distinct spatial patterns. Brain and nervous system cancers showed high-high clustering in Montana, Wyoming, and Nebraska. Ovarian cancer formed a localized high-high cluster in the Pacific Northwest, specifically in Washington, Idaho, and Oregon. Pancreatic cancer was marked by low-low clustering across much of the western United States. Myeloma cancer showed a small high-high cluster in the Southeast, particularly in Virginia and North Carolina. Prostate cancer also displayed a small high-high cluster in the South Atlantic region, specifically in Virginia and West Virginia.

\begin{figure}[t!
]
    \centering
    \includegraphics[width=\linewidth]{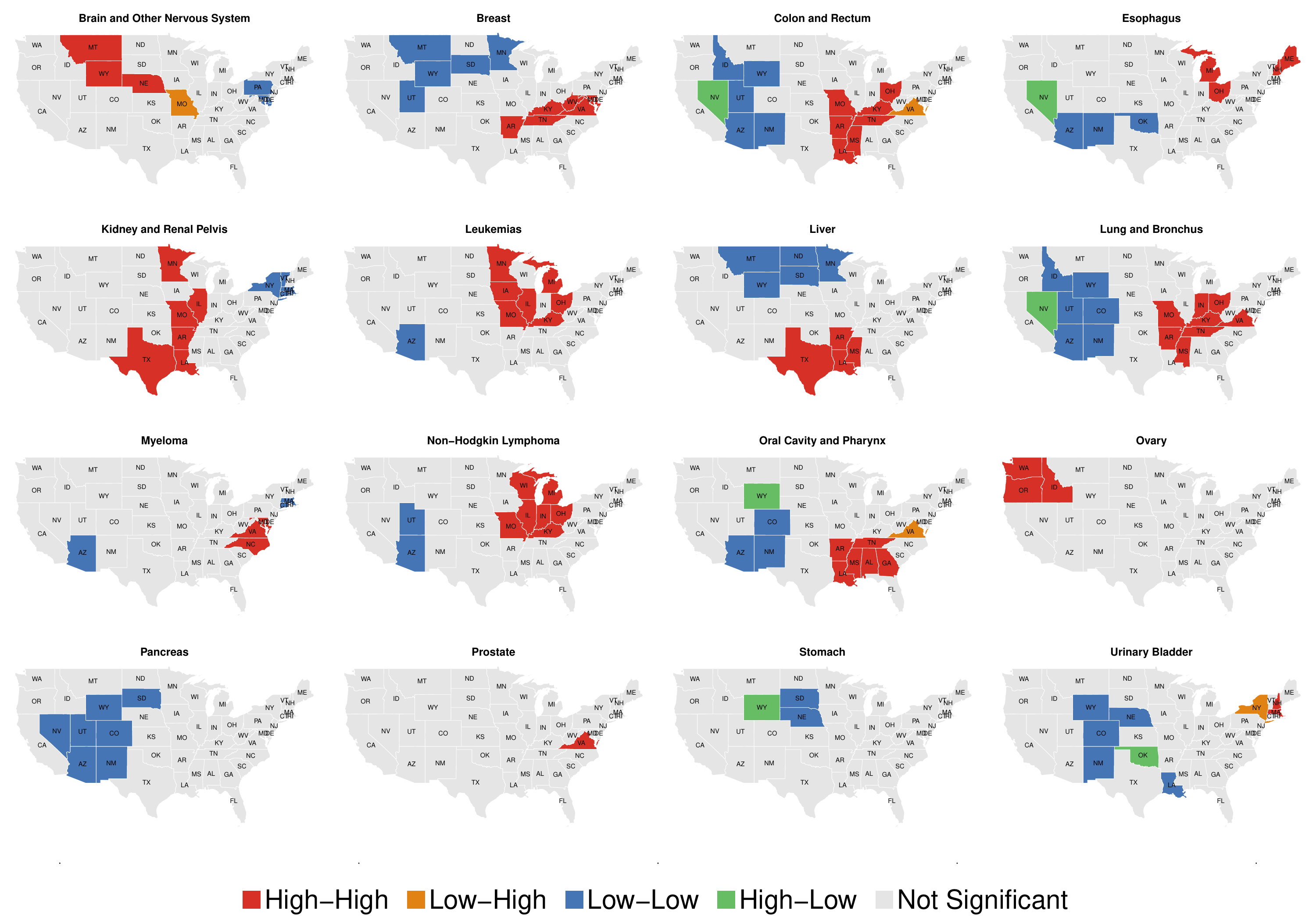}
    \caption{LISA cluster maps of cancer mortality across U.S. states for 16 cancer types, 1999–2021}
    \label{fig:3}
\end{figure}

\noindent Several spatial outliers were detected across some cancer sites. Missouri showed a low-high pattern for brain cancer, while Virginia and New York appeared as low-high outliers for oral cavity and urinary bladder cancers, respectively. Montana and Oklahoma exhibited high-low patterns for oral cavity and urinary bladder cancers, indicating elevated mortality relative to neighboring states. Nevada also showed high-low clustering for esophageal, lung, and colorectal cancers. These patterns, shown in Figure \ref{fig:3}, highlight localized mortality differences across the U.S.

\begin{figure}[htbp!]
    \centering
    \includegraphics[width=\linewidth]{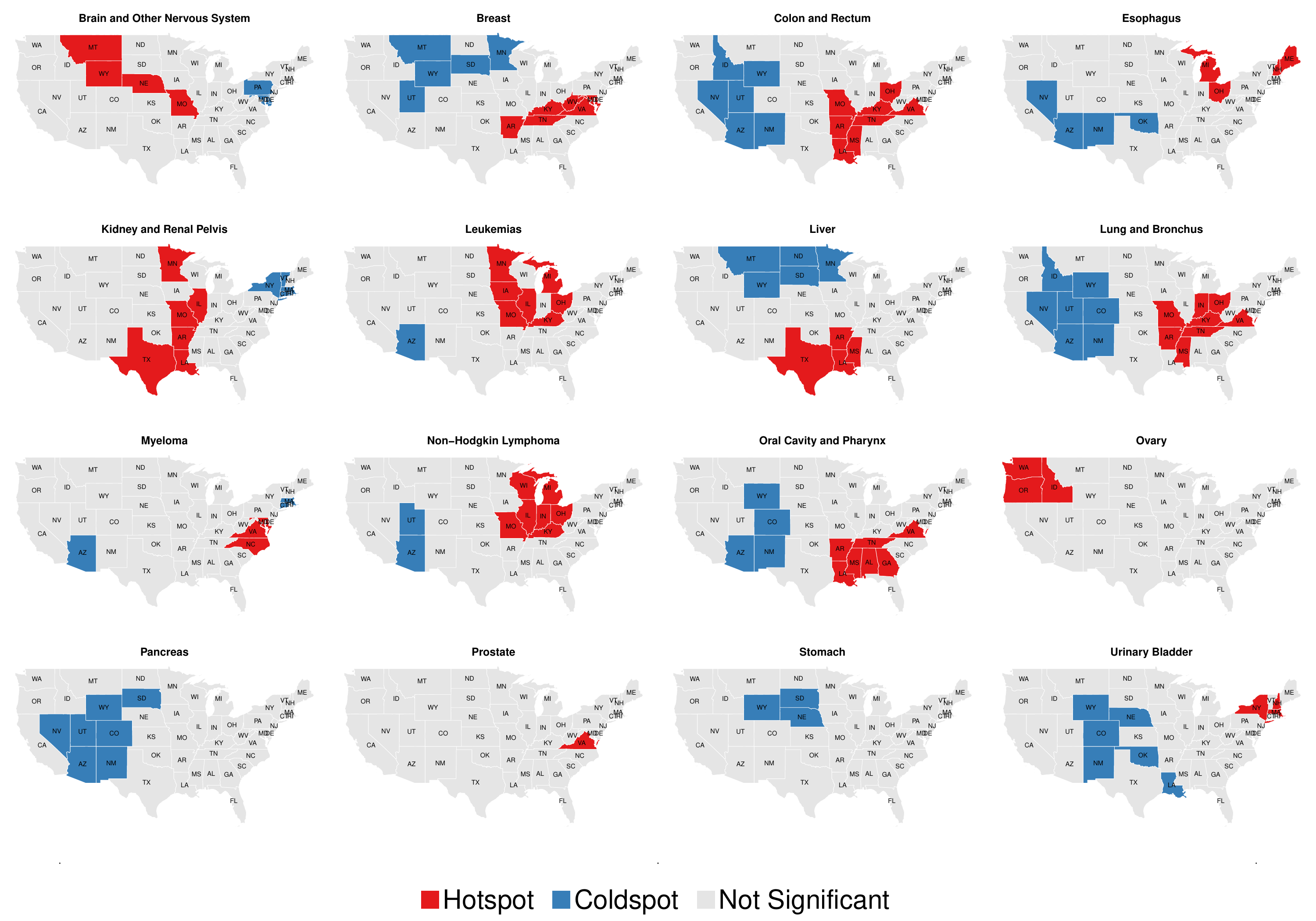}
    \caption{Spatial hotspots and cold spots of cancer mortality across U.S. states, identified using Getis-Ord Gi* statistic (p $<$ 0.05).}
    \label{fig:4}
\end{figure}

\pagebreak
\noindent To further investigate spatial clustering of extreme cancer mortality rates, we applied the Getis-Ord Gi* statistic to identify state-level hotspots and cold spots for each of the 16 consistently reported cancer sites. States with significantly high Gi* z-scores (p $<$ 0.05) were classified as hotspots, while those with significantly low values were designated cold spots. The resulting maps (Figure \ref{fig:4}) revealed spatially coherent clusters, with many hotspots concentrated in the Southeastern and Midwestern U.S., notably for cancers such as kidney and renal pelvis, lung and bronchus, liver, colorectal.

\noindent A frequency analysis of hotspot occurrences across cancer sites identified Virginia, Missouri, and Arkansas as the most frequently flagged states, each appearing in six distinct hotspot maps. These were followed by Ohio and Kentucky (five each), and Tennessee, Mississippi, and Louisiana (four each). This pattern of recurrence suggests that these states bear a disproportionate burden of cancer mortality. Table \ref{tab:hotspot_counts} summarizes the top ten states by the frequency of hotspot appearances across cancer types, highlighting persistent geographic disparities in cancer mortality across the United States.

\begin{table}[htbp!]
\centering
\caption{Top 10 U.S. states with the highest number of cancer mortality hotspots, 1999–2021}
\label{tab:hotspot_counts}
\begin{tabular}{ll}
\toprule
\textbf{State} & \textbf{Hotspot Count} \\
\midrule
Arkansas       & 6 \\
Missouri       & 6 \\
Virginia       & 6 \\
Kentucky       & 5 \\
Ohio           & 5 \\
Louisiana      & 4 \\
Mississippi    & 4 \\
Tennessee      & 4 \\
Illinois       & 3 \\
Michigan       & 3 \\
\bottomrule
\end{tabular}
\end{table}

\pagebreak
\section{\Large Discussion}

\noindent This research advances our understanding of cancer mortality patterns across U.S. states by identifying clusters of states with similar cancer burden profiles. Using hierarchical clustering with Mahalanobis distance, we grouped states based on standardized mortality rates for 16 cancer types from 1999 to 2021. This multivariate approach revealed three distinct clusters with clear geographic boundaries and persistent health disparities.

\noindent The highest burden cluster concentrated in the South and Appalachian regions, showing elevated mortality from lung and bronchus, oral cavity and pharynx, colorectal, and breast cancers. This geographic pattern confirms previous research linking elevated cancer mortality in these areas to entrenched socioeconomic disadvantages, behavioral risk factors, and environmental exposures \parencite{Henley2017,Singh2017}. \textcite{Henley2017} found that rural nonmetropolitan areas had lower cancer incidence rates overall but higher mortality rates than metropolitan areas, particularly for tobacco related and screen preventable cancers. These disparities reflect differences in smoking prevalence, obesity rates, physical inactivity, and barriers to healthcare access, early detection, and timely treatment. In contrast, the low burden cluster included states in the Northeast, Midwest, and West. These states demonstrated consistently lower mortality rates across most cancer types, reflecting better socioeconomic conditions and healthcare infrastructure.

\noindent The application of Mahalanobis distance clustering to our analysis offers a robust framework for identifying states with elevated cancer mortality across multiple sites, revealing co-occurring cancer burdens driven by shared environmental, behavioral, and structural risk factors. This multivariate approach enables the detection of complex population vulnerabilities that are not evident when cancer types are analyzed in isolation \parencite{scott2021geospatial,mccormack2023abstract}. Prior studies have documented how environmental pollutants, socioeconomic deprivation, and social determinants of health contribute to geographic disparities in cancer outcomes \parencite{menon2024cancer,ribeiro2021environmental,schootman2017geospatial}. By incorporating inter-cancer correlations and spatial context, Mahalanobis clustering helps uncover underlying epidemiological patterns that transcend individual disease processes and supports the growing use of geospatial tools in targeted cancer prevention and control strategies \parencite{mccormack2023abstract,schootman2017geospatial}.

\noindent Our spatial analysis confirmed these clustering patterns using multiple statistical approaches. Global Moran's I revealed significant spatial autocorrelation across all 16 cancer sites studied, with particularly strong clustering for kidney, liver, and lung cancers. Local Moran's I and Getis Ord Gi* statistics identified persistent high mortality hotspots in Arkansas, Missouri, and Virginia, where multiple cancer sites showed elevated mortality rates over time. These overlapping hotspots suggest that localized factors create compounded cancer risk in specific regions \parencite{Anselin1995, Moore2017, Bevel2023, Cheng2021}. Previous studies have linked higher cancer mortality to poor healthcare access, lower screening rates, and elevated prevalence of modifiable risk factors like obesity and tobacco use \parencite{belasco2014impact, DeSantis2019}. Breast cancer screening proves especially sensitive to health accessibility barriers \parencite{belasco2014impact}. However, exceptions exist. Alaina et al. (2023) identified counties with low smoking prevalence but high lung cancer rates, particularly among females near the Mississippi River south of St. Louis and among males in Western Mississippi.

\noindent These results suggest that generic, one size fits all strategies may inadequately address region specific risk profiles. Place based interventions such as tobacco cessation campaigns in lung cancer hotspots or enhanced liver cancer screening in high prevalence areas should be guided by these spatial insights. Moreover, linking cancer control programs with broader social determinants of health such as education, food access, and housing may yield greater impact in structurally disadvantaged regions \parencite{karadzhov2024cancer, Anderson2023}. 

\noindent This study has several limitations. State level data may conceal important within state heterogeneity in cancer burden, especially in large or demographically diverse states. While Mahalanobis based clustering accounts for similarity among multiple variables, it does not capture temporal dynamics in mortality rates. Future research should examine county level data, identify temporal trends, and stratify population subgroups to better characterize geographic disparities in cancer outcomes. This would provide more granular insights for targeted interventions and policy development.

\noindent This study demonstrates the value of integrating multivariate clustering and spatial statistics to uncover hidden structures in cancer mortality data. The identification of coherent, geographically concentrated clusters of cancer burden provides a foundation for targeted, equity oriented public health action. By moving beyond single disease approaches, this research offers a more comprehensive understanding of state-level cancer risk in the U.S. that can inform more effective prevention and control strategies.

\section{\Large Conclusion}

This study demonstrates that cancer mortality in the United States is not randomly distributed but follows distinct spatial and multivariate patterns. By combining hierarchical clustering using Mahalanobis distance and spatial autocorrelation measures, we identified coherent groups of states with similar cancer burden profiles and revealed persistent geographic hotspots for multiple cancer types. States in the South and Appalachia emerged as consistently high-burden regions, reflecting underlying structural, behavioral, and environmental disparities.\\

\noindent The findings reinforce the influence of spatial context on cancer outcomes and advocate for localized strategies to mitigate disparities in mortality. Integrating spatial analytics into cancer surveillance can enhance the precision of public health responses and help ensure that interventions reach the populations most at risk. Future efforts should build on this approach by incorporating finer geographic resolution, time trends, and population-level disparities to further inform equitable cancer control policy.

\vspace{-0.5cm}

\section*{Acknowledgment}
\vspace{-0.4cm}
The lead author extends sincere gratitude to all co-authors for their collaborative efforts, intellectual contributions, and support throughout the research process.
\vspace{-0.5cm}
\section*{Conflict of Interest}
\vspace{-0.5cm}
The authors declare no conflict of interest.
\vspace{-0.5cm}

\section*{Funding}
\vspace{-0.5cm}
This research received no specific grant from any funding agency in the public, commercial, or not-for-profit sectors.
\vspace{-0.5cm}

\section*{Author Contributions}
\vspace{-0.5cm}
\noindent \textbf{Author roles:} \\
(1) Research Project: A. Conception, B. Organization, C. Execution \\
(2) Statistical Analysis: A. Design, B. Execution, C. Review and Critique \\
(3) Manuscript: A. Writing of the First Draft, B. Review and Critique \\
\vspace{-0.4cm}

\noindent E.K.: 1A, 1B, 1C, 2A, 2B, 2C, 3A \\
\noindent D.B.: 1A, 1B, 1C, 2A, 2B, 2C, 3B \\
\noindent A.D.: 1A, 1C, 2C, 3B \\
\noindent R.A.: 2A, 2C, 3B \\
\noindent O.J.O.: 1C, 2B, 3B \\
\noindent D.Q.: 1B, 2C, 3B \\
\noindent P.O.A.: 2C, 3B

\printbibliography

\end{document}